\documentclass[fleqn,10pt]{wlscirep}
\usepackage[utf8]{inputenc}
\usepackage[T1]{fontenc}
\title{Speckle tweezers at fluid-fluid interface}

\author[1]{Ramin Jamali}
\author[2]{Sabareesh K. P. Velu}
\author[1,3,*]{Ali-Reza Moradi}
\affil[1]{Department of Physics, Institute for Advanced Studies in Basic Sciences (IASBS), Zanjan 45137-66731, Ira}
\affil[2]{Department of Physics, Rathinam College of Arts and Science, Coimbatore 641021, Tamilnadu, India}
\affil[3]{School of Nano Science, Institute for Research in Fundamental Sciences (IPM), Tehran 19395-5531, Iran}
\affil[*]{moradika@iasbs.ac.ir}

\begin{abstract}
Contemporary approaches to optical multiple micro-manipulation typically involve careful pre-engineering of the laser beam shape. 
In various biomedical and microfluidic scenarios, especially those necessitating unconventional specimen chambers, there is a demand for controlling the collection of micro-objects at fluid-fluid interfaces. 
This requirement arises in contexts such as the transport of materials across liquid interfaces for applications like living cell manipulation, drug delivery, soft functional material creation, and various industrial processes.
For many of these cases, a regular array of trap sites as well as tightly confinement are not essential. 
For such applications at interfaces, we expand on the concept of speckle tweezers (ST), which incorporate randomly distributed light fields for quasi-2D optical manipulation.  
The proposed technique is demonstrated experimentally by applying ST to govern the movement of PS micro-particles at water-oil and water-air interfaces.  The efficacy of the method is validated through the temporal characterization of micro-particle motions using digital video microscopy.
\end{abstract}
\begin{document}

\flushbottom
\maketitle
%
%
\thispagestyle{empty}
\section{Introduction}
Optical tweezers (OT), since their invention in the 1970s, have been used extensively for non-invasive handling of micro- and sub-micro objects, such as cells, organelles, and macromolecules \cite{ashkin1971optical, padgett2011tweezers} to name a few. Therefore, they have become as crucial tools in microbiology and biophysics, for both basic researches and sophisticated applications like optical sorting and optical delivery \cite{macdonald2003microfluidic}. 
Through programmable spatial redistribution of light multi-object systems for specific applications may be obtained. For example,   a lattice of micro-particles can be formed using periodic optical potentials, i.e. multiple traps \cite{albaladejo2009giant}. 
Extending a single OT into multiple OTs which is also called holographic optical tweezers (HOT) has been an exciting advancements of OT and led to numerous applications \cite{grier2003revolution}.  The simple case of a dual OT can be readily performed by utilizing separate steering mirrors and a pair of beam splitters \cite{misawa1992multibeam}. 
HOT can be produced by time sharing of a single trapping laser beam through the use of different steering devices,	such as a rapid scanning galvo-mirror or an acousto-optic beam deflector (AOD) \cite{visscher1996construction,emiliani2004multi}. 
However, none of these scanning approaches allow simultaneous and easy implementation of the trapped particles positions, which is a crucial requirement in most of the applications. The inverse problem of diffraction phenomenon, i.e., designing a diffracting element that generates a pre-designed on-demand wavefront via diffraction of an illuminating laser beam. These elements are called diffractive optical elements (DOEs) and they can be designed by the use of iterative algorithms on the basis of phase retrieval from diffraction formula \cite{soifer2002methods,cojoc2004multiple}.  Fabrication of DOEs can be done by lithography techniques which provide very precise 3D  structures that diffract laser beams in reflection to generate the pre-designed beam shapes in the diffraction plane \cite{cojoc2005laser}.  However, this approach is expensive and cannot be appropriate for dynamic phenomena as the use of successive elements to refresh the generated wavefront requires the fabrication of multiple DOEs for a simple dynamic phenomenon and their application at high frame rates is not possible.  
The advancement of spatial light modulators (SLMs) that act as programmable DOEs to engineer a single laser beam wavefront toward splitting it into many optical trapping beams. These trap sites can be independently controlled in terms of strength, position in 3D, and shape at the frame rates that the SLM refreshment rate allows \cite{grier2003revolution}. The HOTs via the use of SLMs have found numerous applications in biology, statistical mechanics, fluidics, etc. for fundamental and applied researches \cite{volpe2023roadmap}.  The first commercial apparatus for dynamic HOTs was implemented by  Arryx, Inc. and received the R\&D 100 Award for Technical Innovation in 2002 \cite{DavidG2002}. The use of DOEs also enables the generation of further structured beams,  such as  Laguerre-Gaussian beams, Bessel beams, Mathieu beams, etc., which have found the wider spectrum of applications in the aforementioned fields \cite{simpson1996optical,arlt2001optical,alpmann2010mathieu}. Specifically using the central part of  Laguerre-Gaussian or Bessel beams have been used to trap microscopic objects with refractive indices lower than their surrounding medium and metallic particles \cite{rui2015manipulation}. These objects cannot be trapped with commonly used  Gaussian beams in OT as their negative or imaginary refractive index causes repelling them from the trap site, which is avoided by the use of the beams with central dark regions.  For multiple trapping of such samples, similarly, the generation of arrays of structured beams is the common solution \cite{grier2003revolution}. 

Nevertheless, for numerous applications that require simultaneous manipulation of micro-objects, the use of HOT using the aforementioned methods is more than sufficient in two terms:  (1) the regular arrays which need pre-design efforts are not always demanded; (2) brief confinement of a set of micro-objects is sufficient and stiff OTs are not always required. 
For example, for guiding or sorting of a mixture of two micro-particles of different sizes in a fluidic channel, one only needs to keep one  size slightly longer than the other size to sort them while there are flowing.  
For this example and many other ones, speckle tweezers (ST), elegantly fulfills the above requirements.
To show that we first explain the speckle fields. Upon interference of several light waves that are originated from a coherent source but with random phases or amplitudes, an intensity distribution is formed which is called the speckle field. The speckle field can be generated by illuminating a reflective rough surface,  passing the light through an inhomogeneous medium, random mixing of propagating modes in a multimode fiber, or by intentionally addressing properly designed DOEs onto a SLM \cite{goodman1976some,goodman2007speckle}. In many optical imaging systems the formation of speckle is avoided toward a noise-free image by means of coherence-decreasing methods, such as a rotating diffuser \cite{milkowski2009speckle}. However, in recent years it is proved that this phenomenon should be welcomed as statistical processing on its distribution may provide valuable information about the generation mechanism that generated the field \cite{bender2018customizing,potenza2010measure,pincce2016disorder}. Therefore, analyses based on speckle fields have led to novel imaging, detection, and measurement methodologies: dynamic laser speckle analysis has been used to study scaffold activity, characterization of biodegradable nanocomposites, and detection of alignment in multicomponent lipids, microstructure movements, pitting corrosion, conduction mechanism in memristor device  \cite{rad2020non,jamali2023surface,panahi2022detection,pedram2023evaluation, jamali2024speckle}. The secondary speckle analysis provides easy-to-implement yet precise measurements, for example, multiple speech sources and heart beats \cite{zalevsky2009simultaneous}, speckle rheology provides a non-contact and non-destructive rheological parameters measurement \cite{hajjarian2020tutorial}, and imaging based on contrast parameter of speckle fields is becoming an established imaging method for the cases of highly diffusive samples \cite{senarathna2013laser}. Nevertheless, in this research, we overlook what has generated the speckle pattern, and we use the intensity distribution of speckle patterns for trapping.
Since the speckle patterns contain bright grains surrounded by a network of dark regions,  there are lots of local intensity gradients throughout the speckle pattern, and each of them can act as an OT. Therefore, a speckle pattern can be used as an irregular array of optical traps, and this is called speckle tweezers (ST) \cite{volpe2014brownian,jamali2021speckle}. 
More interestingly, as there are positive and negative intensity gradients throughout a speckle pattern, ST can be used to simultaneously manipulate any type of micro-objects, including dielectric particles with refractive indices higher than surrounding medium, e.g., polystyrene (PS)  in water, or lower than surrounding, e.g., bubble in water, metallic micro-objects, etc.  \cite{jamali2021speckle}.
Furthermore, collective motion control by ST due to their intrinsic randomness may not be a disturbing factor in the manipulation of a set of particles, and even it can be useful as it uniformize the influence of the light field on the particles' motions. 

In the first demonstration of ST, V. Shvedov et al. have shown that high-intensity volumetric speckle light fields can selectively trap light-absorbing micro-particles in a gas \cite{shvedov2010selective}.  Beforehand, periodic potentials had been extensively studied to produce light-absorbing effects and for particle guiding and sorting \cite{pesce2020optical}.
The stochastic motion of particles in random potentials is the foundation of many phenomena, from the molecules undergoing anomalous diffusion within the cytoplasm of a cell to the Brownian motion of stars in galaxies  \cite{pesce2020optical}. F. Evers et al. showed that the appearance of anomalous diffusion in colloids is correlated to the applied optical speckle fields \cite{evers2013colloids}. 
The field of active matter, which beside the Brownian motion possesses active noises,  also is a grown-up field and specifically required the use of light-to-work conversion \cite{rubinsztein2016roadmap}. The ST field includes a random, but somehow tunable, distribution of isolated high-intensity regions surrounded by irregular and interconnected dark regions. The dark spots are, indeed, much more common than bright ones \cite{gateau2019topological}. 
In this regard, the nature of the speckle fields makes them as an elegant possibility for collective manipulation of mixtures of low and high refractive index micro-particles, such as trapping, guiding, and sorting, and, therefore, a potential method for targeted drug delivery \cite{volpe2014brownian,jamali2021speckle,sadri2024sorting}. One of the first experiments on the use of ST in this range demonstrated the emergence of anomalous diffusion in colloids and showed the control of the motion of Brownian particles \cite{chubynsky2014diffusing}. The optical speckle field has been also used for producing a thermal speckle field through interaction with plasmonic substrates and converting the high-intensity speckle grains into the corresponding thermal speckle grains  \cite{kotnala2020opto}.  In another study,  it has been shown that the density of the structural defects in a 2D binary colloidal crystal can be engineered using a speckle field \cite{nunes2020ordering}. Based on a memory equation a theoretical model has been suggested explanation of the motion of colloidal particles within the ST field \cite{pincce2016disorder}. Furthermore, by simulation of the force field throughout the ST field, the trajectories of microscopic particles under ST are derived \cite{garcia2018high,jamali2021speckle,volpe2014speckle}.  

In this research, we expand on the concept of ST for controlling of the collective motion-related phenomena at fluid-fluid interfaces. 
Interfaces are the meeting points of different phases of matter, hence, they are omnipresent in physics, biology, chemistry, and engineering \cite{binks2017colloidal}.  Among different interfaces water-oil and water-air interfaces play crucial roles in a wide spectrum of phenomena, such as oil spill remediation, enhanced oil recovery, water treatment, purification etc. \cite{hou2022review,wen2020evidence,colloids3020050,deal2021water,baret2012surfactants,kortmann2009rapid,nikiforidis2013purified}.  
The water-oil interface is a region where two immiscible liquids meet and interact. The interaction is governed by interfacial tension and the interface is characterized by intriguing molecular interactions \cite{binks2017colloidal}. Interfacial tension is a measure of the force that keeps the two liquids separate and arises due to the imbalance of cohesive forces at the interface. In the bulk of a single liquid, molecules are surrounded by similar molecules and experience balanced attractive forces. However, at the interface, molecules are in contact with different types of molecules and experience unbalanced forces leading to interfacial tension \cite{binks2017colloidal}. In the case of the water-oil interface, water is a polar molecule, meaning it has a positive and a negative end, which promotes hydrogen bonding among water molecules. Oil, on the other hand, is non-polar and does not have a positive or negative end \cite{silverstein1998real}. This difference in polarity leads the two liquids to layer one on top of the other at the macroscopic scale. 
Therefore, adding surfactants, which are molecules with both hydrophilic and hydrophobic parts, can reduce the interfacial tension. They align themselves at the interface, with their hydrophilic heads in the water and their hydrophobic tails in the oil, which in turn, reduces the direct contact between water and oil molecules, thereby decreasing the interfacial tension and increasing the miscibility of the two liquids \cite{binks2017colloidal}.
The water-air interface, on the other hand, is a physically and chemically active environment, and plays a crucial role in phenomena as diverse as climate modeling and optical refraction \cite{deal2021water}. 
Some everyday phenomena, such as the spherical shape of small droplets and the ability of some insects to walk on water are understood through consideration of the water-air interface. 
Additionally, the refraction of light at the water-air interface, caused by the change in speed of light as it moves from water to air, is responsible for optical phenomena such as the bending of a straw in a glass of water \cite{liss1988water}.

Understanding these interfaces is not just a matter of scientific curiosity. It has practical implications in various fields. For instance, the study of water-oil interfaces can enhance techniques in oil recovery and environmental cleanup \cite{hou2022review}. Insights into the water-air interface can inform climate models, as this interface plays a crucial role in the exchange of gases between the atmosphere and the oceans \cite{tang2020molecular}. 
However, the interfaces between different phases of matter are complex and dynamic regions, which offer a rich area of study with far-reaching implications in science and industry, from oil recovery to atmospheric investigations.

Most studies on the dynamics of microswimmers in inhomogeneous multi-phase systems have focused on swimming in the vicinity of boundaries, mainly solid-fluid boundaries and fluid-air interfaces \cite{li2014hydrodynamic}.
For example, transmembrane transport of bacteria and viruses is a key stage in infection. 
Lauga et al. showed a clockwise circular swimming motion of E. coli near a solid-fluid boundary  \cite{lauga2006swimming}, and Leonardo et al. reported a counterclockwise rotation near a fluid-air interface \cite{di2011swimming}. 
Another surface-active example is a microgel. Microgels absorb at the fluid interface covering the maximum fraction of the area available to minimize the contact line between the two immiscible fluids and reduce the interfacial tension, which leads to the microgel flattening at the interface \cite{plamper2017functional}.  
Research into the dynamics of swimmers in such environments is of vital importance to understand biological transport and development of artificial micro-machines to be used for targeted delivery in complex fluid environments \cite{yan2015magnetite,laage2006molecular}. In \cite{de2022brownian} explores the experimental study of the motion of micrometer-sized water droplets, formed spontaneously at the interface between water and oil, on an out-of-thermal equilibrium surface. Despite the dynamic nature of the process and the resulting system heterogeneities, the droplet motion exhibits a Brownian component and a drift component due to gravity, without any observed anomalous diffusion.

Many of the aforementioned applications include trapping at interfaces. For instance, Masuhara et al. investigated dense-liquid droplet formation and the crystallization of organic molecules, and  Lida et al. demonstrated induce of hybridization by optical assembly of DNA-modified nanoparticles at water-air interfaces \cite{iida2016submillimetre}. 
Also,  rotational dynamics of nanoparticles under a circularly polarized optical vortex due to light-induced interactions at an interface is predicted \cite{tamura2019dynamics}.  Recently, in \cite{kim2024optical} the actuation of the liquid-liquid interface by illumination of laser on nano-particle loaded particles toward interfacial material manufacturing and reconfigurable structures has been demonstrated. 

ST is the incorporation of another feature of light, which is the transfer of optical momentum to microscopic particles via the creation of multiple intensity gradients throughout a surface containing the particles. It seems that the quasi-2D nature of ST fits well with the geometry of fluid-fluid interfaces. Therefore, ST offers an elegant possibility to control the motion of particles at the interfaces, which is addressed in this paper.

The paper is structured as follows. Section \ref{experiment} details the experimental procedure and setup. Section \ref{analysis} explains the data analysis procedure. Section \ref{theoretical} describes the theoretical model for the fluid-fluid interface. Section \ref{results} presents and discusses the findings. Finally, the paper is concluded in Section \ref{conc}. 

\begin{figure*}[t]
	\begin{center}
		\includegraphics[width=0.95\linewidth]{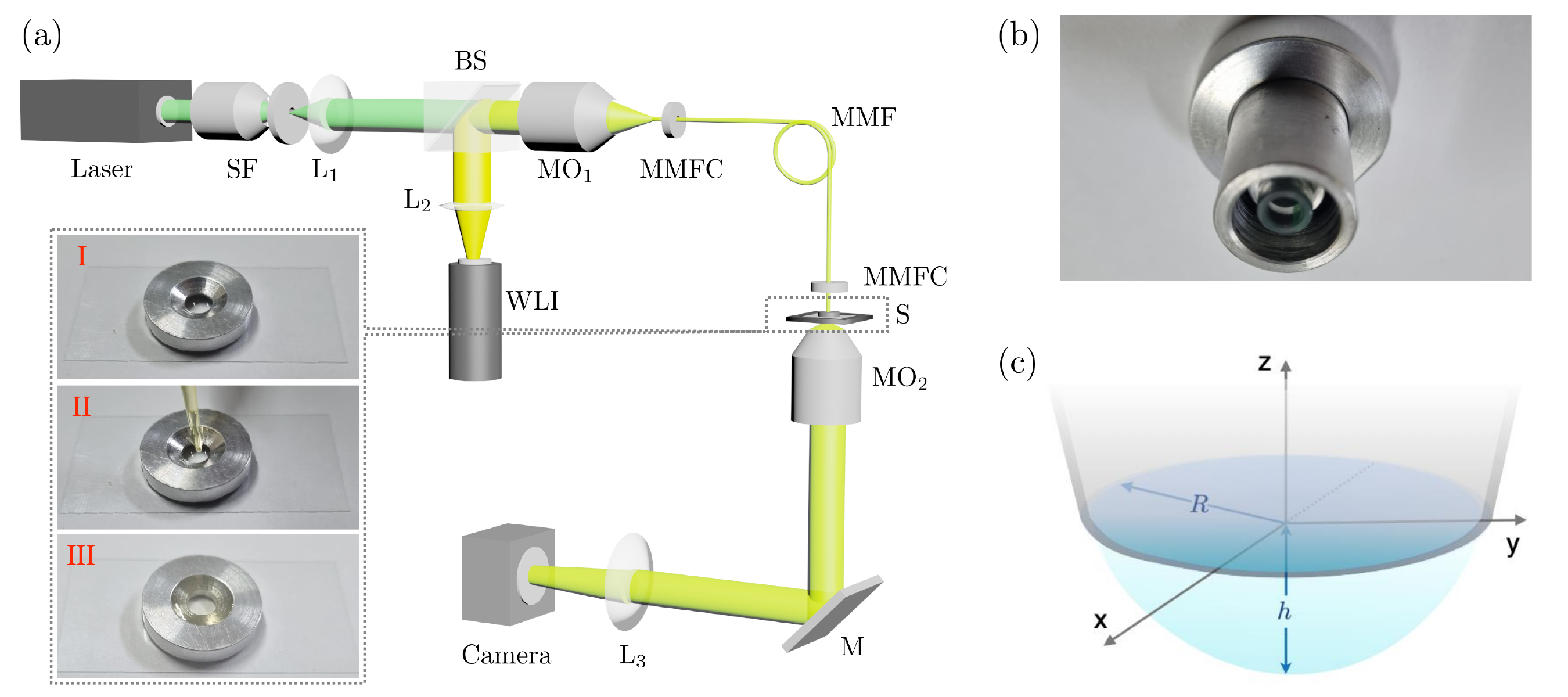}
		\caption{(a) Schematic of the home-built experimental setup; SF: spatial filter, MO: microscope objective, MMF: multi-mode fiber, MMFC: multi-mode fiber coupler,  BS: beam splitter, L: lens, M: mirror, WLI: white light illumination, S: sample. Inset: Enlarged view of the sample chamber used for water-oil interference experiment, and the three stages, (I)  containing only distilled water; (II) oil injection into the chamber; (III) containing water and oil, and forming the interface. (b) Picture of the sample chamber used for water-air interface experiment, through the so-called hanging droplet arrangement. (c) 3D schematic of the hanging droplet and the curved liquid-air interface. $h$ and $R$ represent the height and radius of the hanging droplet, respectively. The inner radius of the tube is also $R=$ 1 mm.}
		\label{Figure01}
	\end{center}
\end{figure*}

\section{Experimental procedure and methods}
\label{experiment}
Figure \ref{Figure01}(a) shows the schematic of the experimental setup that we used for speckle trapping of collections of micro-particle at the water-oil and water-air interfaces. The setup consists of three parts: (1)  apparatus for controlled speckle field generation, (2)  home-made optical microscopy, and (3) a specific sample chamber for each of the interfaces. 
The optical microscopy and ST are integrated so that the microscopy illumination light and the ST beam are common-path. 
A solid-state laser beam (Shafa Parto Parse CO, PSU-FC, 532 nm) passes through a spatial filter (SF), which filters out the unwanted spatial frequencies in the Fourier space through placing a pinhole in the focal plane of a lens, and is collimated by the use of another lens (L$_1$, focal length = 50 mm) to provide a uniform beam profile. The collimated beam passes through a beam splitter (BS), which is used to combine microscopy illumination light, and, by the use of a microscope objective (MO$_1$, Olympus, 10 $\times$, plan achromat, NA = 0.25, WD = 10.6 mm),  is focused onto the input entrance of a multi-mode fiber (MMF) mounted on a kinematic fiber mount (MMFC). 
The MMF has a numerical aperture of NA = 0.22, a core diameter of $365 \pm 14$ $\mu$m, and refractive indices of 1.4589 and 1.4422 for the core and cladding, respectively.
According to these parameters up to about 474 laser beam modes can propagate through the MMF. These modes are superposed, and after a short propagation, the interference of these modes with random phases forms the speckle pattern at the output of the fiber which illuminates the sample (S) to build STs. 
Generation of speckle patterns by the use of MMF has considerable advantages over rough surface or diffusive translucent material, such as the possibility to form uniform speckle fields over programmable regions, adaptability, and portability in the device's design, and improved transmission efficiency. Moreover, by adjusting the sample-MMF distance the average grain size of the speckle field may be tuned. 
The conventional inverted microscope is combined to the ST setup to image the sample under ST. A diverging light of a white light source (WLI) is collimated by L$_2$ (focal length = 100 mm) and is redirected and focused onto the input of the MMF by BS and MO$_1$, respectively. The microscope objective MO$_2$ (Nikon, $10\times$, CFI plan achromat, NA = 0.25, WD = 10.50 mm), mirror (M), and tube lens (L$_3$, focal length = 50 mm) form a microscopic image on the digital camera (Thorlabs, DCC1645C,  1.3 Megapixels (1280 $\times$ 1024), 8-bit dynamic range, 3.6 $\mu$m square pixel pitch). 

The inset of Fig. \ref{Figure01}(a) shows the specific chamber for water-oil experiments and the procedure to form an interface and to place the micro-particles at the interface.   The chamber is a cylinder of 4 mm thickness and 20 mm diameter made of 7000 series Aluminium. Up to 2 mm of the height of the cylinder, in the bottom has an opening of 5 mm (inner diameter of the cylinder) and the opening of the upper surface has a 10 mm diameter. Therefore, the inner part is a conical frustum with an angle of 51.3$^{\circ}$. The whole chamber is glued on a coverslip and sealed. 
For the water-oil experiments, the lower part is filled with distilled water. Then, 1 $\mu$L of diluted 1 $\mu$m PS micro-particles are injected by careful placing of an insulin syringe needle on the slant, and connecting the tip to a fluidic injection pump (FNM Co., SP102 HSM) via a fluidic tube to ensure a quiescent placement of PS in proximity of the interface and to avoid disturbances at the interface. Finally, oil ($\eta_{\rm{o}}$ = 0.122 Pa.s, $\rho_{\rm{o}}$ = 970 $\frac{\rm{kg}}{\rm{m}^3}$) is infused into the chamber to fill up the conical frustum part and forms the interface. 
 
For the water-air interface experiment, another specific chamber is designed and fabricated, which is based on the use of hanging droplet formation. The chamber includes three detachable parts (Fig. \ref{Figure01}(b)). Part I is a mountable post to hold the chamber in the setup. The outer part (II) is a cylinder of 22 mm height, 16 mm outer diameter, and 12 mm inner diameter. The inner part (III) is a hallo cone with a cylinder-shaped base. The base has a 3 mm height, 15.5 mm outer diameter, and 6 mm inner diameter. Therefore, the part II can be placed and fitted on the part II stably. The halo cone has an inner cylindrical shape of 6 mm diameter. Its height is 17 mm from the base, and the outer diameter near the base is 11.5 mm and at the end is 9.5 mm, making a cone with 3.4$^{\circ}$ slope angle.  These geometrical specifications provide a feasible platform to prepare a hanging droplet. In the experiments, part II is placed on a coverslip and actually plays the role of a holder, and a plastic or glass pipette tip ($R \approx$ 1 mm) is placed in the halo cone and filled up with water with a hanging droplet at the free end. According to the designed specifications, the droplet can not reach the surface of the glass slide.  The height of the droplet $h$ is a function of its volume, therefore by adjusting the volume, it can be controlled. For example, for the above specifications, an 80 $\mu$L droplet has an approximately 1 mm height. The details of the geometry of a hanging droplet is discussed in  \cite{savino2004transient}.   
A hanging droplet geometry is schematically shown in Fig. \ref{Figure01}(c).  
\begin{figure*}[t]
	\begin{center}
		\includegraphics[width=0.95\linewidth]{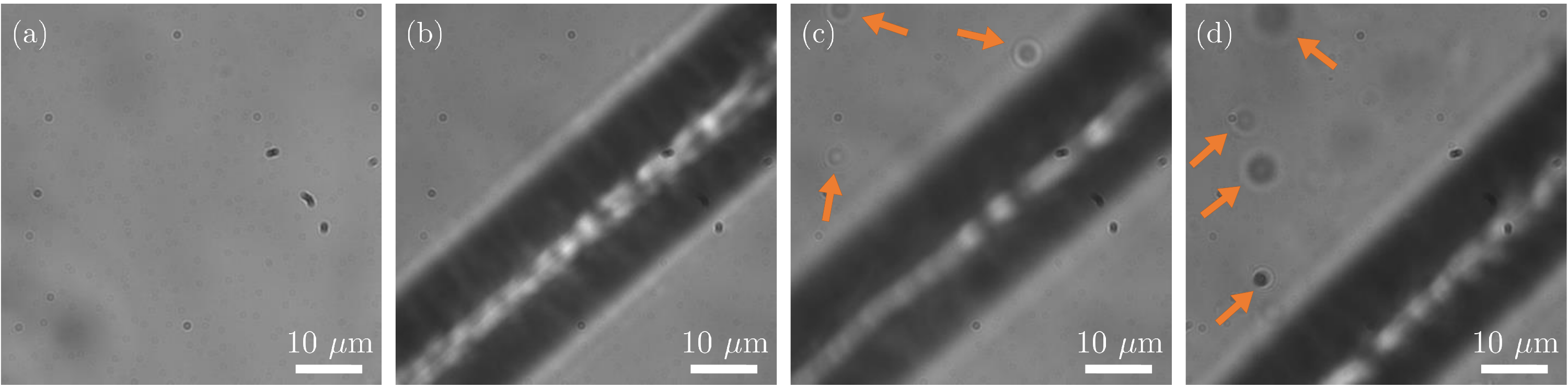}
		\caption{Bright field images of the step-by-step formation of the interface between water and oil and injection of PS micro-particles. (a) The chamber is filled up with water only; (b) A tracer is added to the surface of water and properly imaged; (c) The micro-particles are injected to the same depth of the tracer. (d) Oil is infused and the particles at the same sharpness of the tracer are subjected to follow-up tracking. For the further experiments, the imaging configuration is remained unchanged.   The arrows indicate some of the micro-particles at the interface between two liquids.}
		\label{Figure02}
	\end{center}
\end{figure*}

In both cases, i.e., water-oil and water-air interfaces, the corresponding chamber is placed on the stage of the home-made microscope. It is obvious that with conventional microscopy finding and imaging the interface, especially in the water-oil case, is a challenge.
In order to overcome the issue, we perform an auxiliary experiment in the beginning, in which only the lower part of the chamber is filled up with water. Then, a tiny object is carefully placed on the free surface as a tracer, and the imaging camera and the lens L$_3$ distances are adjusted so that a sharp image is obtained. This imaged plane, therefore, represents the interface even after adding oil as the imaging apparatus is placed below the chamber. Upon finding the interface plane the PS micro-particles are injected and then oil is added. In Fig. \ref{Figure02}  step-by-step formation and finding the interface between water and oil and injection of micro-particles are shown. 

In the case of the water-air interface, the particles are gathered at the interface of the hanging droplet due to the gravitation. However, they can move almost freely in 2D. In the case of the water-oil interface the particles injected to the quasi-2D region at the interface, can fall also after a while beside their lateral motions. We show that ST is capable to restrict particles at the interfaces.  
The primary deterministic confining forces operating on dielectric particles with sizes comparable to or smaller than the typical speckle grain size are the optical gradient forces, which drag particles with a high refractive index toward the optical field's intensity maxima.
\section{Theoretical model}
\label{theoretical}
The intricate motion of micro-particles at the interface of two immiscible fluids, like water and oil or water and air, has been intensively interested by scientists for decades. The micro-particles, which can be organic or inorganic, have the potential to alter the properties of the interface, leading to unique interfacial phenomena. There are several fascinating and complex phenomena which include micro-particle motion at the fluid-fluid interface and have significant implications in various fields, including  Pickering emulsions, self-assembly, flotation, encapsulation, drug delivery, microrheology, microfluidics, material science, environmental science, and biomedical engineering \cite{villa2020motion}.  The motion of the micro-particles at interfaces is influenced by various factors including viscous drags, fluctuations of the fluid interface, and its physicochemical properties. For example, in the case of a water-oil interface, micro-particles can act as surfactants, reducing the surface tension and stabilizing the interface \cite{baret2012surfactants,abras2012mobilities}. This has important applications in emulsion stabilization, where the prevention of coalescence is crucial. Also, at the water-air interface, micro-particles can form a layer, creating a barrier that prevents evaporation and gas exchange. This is particularly relevant in the context of environmental pollution, where micro-particles can significantly impact the gas exchange between water bodies and the atmosphere  \cite{barboza2019macroplastics}. 

Confined fluids encounter boundaries, such as solid walls or interfaces with other fluids. These boundaries can exhibit either stick or slip behavior, impacting the fluid's motion relative to the boundary. A ``stick'' condition occurs when the fluid has zero relative velocity at the contact point. Conversely, a ``slip'' condition occurs when there is relative motion at the contact point. 
Under laminar flow near a solid, flat surface, Newtonian fluids are generally assumed to exhibit a stick boundary condition. This implies that the fluid velocity matches the solid surface's velocity, leading to an increased drag force. Consequently, a simple velocity profile across the fluid flow shows a minimum value at the solid-liquid contact, increasing towards the center of the flow.

To elucidate the underlying physical principles governing the particle motion at interfaces physical concepts and mathematical models have to be considered.   
By understanding the forces acting on the particles and the resulting dynamics, one can gain insights into the aforementioned governing mechanisms. The behavior of particles at fluid-fluid interfaces is essentially a complex interplay of forces, disruptions, and boundary conditions. In the context of an unrestricted fluid, an interface disrupts the uniformity and consistency of the space.
This disruption becomes evident in the form of a tensorial and spatially variable micro-particle drag when a micro-particle moves near the interfaces. The values of this micro-particle drag are dictated by the boundary conditions at the interfaces. The well-recognized theoretical formulas for translational mobilities can be applied to the particle motion, both parallel and perpendicular to an interface \cite{villa2020motion}. 
H. Brenner achieved a significant breakthrough by deriving the translational mobility for micro-particles moving at a right angle to an infinite planar interface \cite{brenner1961slow}.  
The continuity of shear stress at such interfaces is a crucial boundary condition.
H. Brenner computed the complete series solution for the quasi-static Navier-Stokes equations, in both cases of full-slip and no-slip boundary conditions  \cite{brenner1961slow}. 
S. H. Lee et al. theoretically investigated the Navier-Stokes equations for a spherical particle immersed in fluid approaching a flat fluid-fluid interface under the assumption of creeping flow \cite{lee1979motion}.  Their study focused on the disturbance generated by the particle, and, indeed, was the extension of the Stokeslet solution and stream functions provided earlier by H. A. Lorentz \cite{lorentz1907abhandlungen}.

Obtaining an analytical solution for the full degree-of-freedom motion of particles near the interface, however, is more challenging due to the interplay between translational and rotational movements. Moreover,  when a parallel translational force is applied to a finite-sized object,  the drag on the object depends on  its distance to the interface  and results in a non-zero torque. For the fluid-fluid interfaces, on the other hand,  the maximum velocity in the profile occurs at the free interface. 

For the motion of micro-particle at the interface we consider the four major influencing factors of interfacial tension ($\gamma$), particle-fluid interaction, buoyancy force ($F_b$), and the hydrodynamic forces.
The interfacial tension factor arises due to the unbalanced forces at the molecular level within the fluids. It acts like a stretched membrane at the interface resisting the expansion of the interface area. Particle-fluid interactions depend on the particle's surface level of hydrophilicity/hydrophobicity and the surrounding fluids. Hydrophobic particles tend to minimize contact with water and favor the oil or air phase.
Buoyancy force depends on the particle volume ($V_p$), density ($\rho_p$), and the density difference between the micro-particle and the supporting fluid ($\Delta \rho$), and is calculated from  $F_b = V_p g \Delta \rho$, where $g$ is the gravitational acceleration. ``Supporting fluid'' refers to the denser fluid that the micro-particle is partially submerged in.

The hydrodynamic forces include drag force ($F_d$), due to the relative motion of the micro-particle, and the fluid and lift force ($F_l$), arising from the flow gradients near the interface.
Lift is the force acting on an object moving through a fluid and is perpendicular to the direction of the fluid flow. This force arises due to the pressure difference between the upper and lower surfaces of the object, not due to the pressure or stress themselves. For flows with minimal disruption, the impact of inertial forces becomes negligible, allowing the system to be described by the framework of Stokes flow. In this regime, the fluid behaves as if it primarily resists motion through viscous dissipation. Fluids exert forces on immersed objects due to interactions at the interface. 

The interplay between the aforementioned forces for the micro-particle motion at the interface may be described using Newton's second law of motion and the net force acting on the micro-particle:
\begin{equation}
	F_{net} = F_{b} + F_{d} + F_{l} - \gamma L,
	\label{EqNet}
\end{equation}
where $L$ represents the length of the contact line between the micro-particle and the interface. The negative sign for the interfacial tension term indicates its opposition to the expansion of the interface area.
The interfacial tension contribution depends on the particle's shape and the contact angle ($\theta$) with the interface. For a spherical particle of radius $a$ which is partially submerged at the interface is $L = 2\pi a  \cos\theta$, where $\theta = 0$ for a perfectly wetting particle (completely submerged in water) and $\theta = 180^{\circ}$ for a completely non-wetting particle (on the oil or air side) \cite{pritchard2016fox}.
The buoyancy force pushes the particle toward the denser phase and is defined as $F_b = \frac{4}{3}\pi a^3 g \Delta \rho$. Drag and lift forces can be approximated with $F_d = 6 \pi a v \eta$, where $\eta$ is the fluid's viscosity and $v$ is the particle's relative velocity \cite{pritchard2016fox}. 

By solving the equation Eq.~\ref{EqNet}   the particle's motion under different conditions can be analyzed. However, due to the non-linear nature of the drag and lift force terms, analytical solutions might not always be possible. Indeed, considering the complexity of analytical studies on particle behavior at interfaces, mostly numerical approaches have been followed. For example, A. J. Goldman et. al presented the first numerical solutions to the Stokes equation for a sphere moving parallel to a solid wall  \cite{goldman1967slow}, and A. V.  Nguyen and G. M. Evans identified the exact numerical solutions corresponding to full slip boundary conditions on the plane \cite{nguyen2004exact}. At a water-air interface, the particle drag is anticipated to be controlled by the free-slip boundary condition, which is consistent with predictions for unconfined fluid surfaces. The free-slip boundary significantly influences the motion of the micro-particle near the interface, and considering this condition is a reasonable assumption for the water-air interface due to the negligible shear stress at the interface \cite{pritchard2016fox}.

Based on considering the aforementioned involving factors, the resulted theoretical framework may provide a foundation for understanding the motion and potential confinement of micro-particles at fluid-fluid interfaces in general cases. We employ a model that combines the findings from Perkins and Jones (PJ) regarding free interfaces with the second-order adjustments, proposed by Lee et al.  \cite{perkins1991hydrodynamic,lee1979motion}. This model derives a reflection theorem allowing to solve the linear (quasi-static) Navier-Stokes equations \cite{perkins1991hydrodynamic}.
It is shown that  PJ's results exhibit favorable behavior due to the removal of singularities near the interface at small distances \cite{benavides2016brownian}. Consequently, this model appears more appealing than the one introduced by Brenner   \cite{brenner1961slow}. For a spherical particle with a radius $a$, positioned above a fluid-fluid interface at a distance $h$, the normalized friction coefficient  in a parallel direction, $\frac{\xi_\parallel^{PL}(h)}{\xi_0}$, depends on the viscosity ratio, $\lambda = \frac{\eta_1}{\eta_2}$,   and can be expressed as \cite{benavides2016brownian}:
\begin{equation}
	\frac{\xi_\parallel^{{\rm{PL}}}(h)}{\xi_0}  = \frac{\digamma_0}{\digamma_\parallel^{{\rm{PL}}} (h)} =  	\frac{\xi_\parallel^{{\rm{PJ}}}(h)}{\xi_0} + \frac{15\lambda}{16(1+\lambda)} \{(\frac{a}{h}) + \frac{3\lambda - 12}{16(1+\lambda)} (\frac{a}{h})^2\},
	\label{eq4par}
\end{equation}
where $\xi_0$ and $\digamma_0$ are the drag and diffusion coefficients of the particle, respectively. In equation Eq. \ref{eq4par}, $\xi_{\parallel} ^{{\rm{PJ}}}(h)$ is the distance-dependent friction coefficient of a soft interface for the parallel direction, obtained within the PJ approximation.
The resulting friction and diffusion coefficients are denoted as PL to address the combination of PJ approximation and Lee model. $\digamma_\parallel^{{\rm{PL}}} (h)$ is the distance-dependent diffusion coefficient along the direction parallel to the free interface. 
Similarly, the behavior of  the spherical particle in the perpendicular direction may be described by:
\begin{equation}
	\frac{\xi_\perp^{{\rm{PL}}}(h)}{\xi_0}  = \frac{\digamma_0}{\digamma_\perp^{{\rm{PL}}} (h)} =  	\frac{\xi_\perp^{{\rm{PJ}}}(h)}{\xi_0} + \frac{3\lambda}{8(1+\lambda)} \{(\frac{a}{h}) + \frac{15\lambda + 12}{8(1+\lambda)} (\frac{a}{h})^2\}.
	\label{eq5per}
\end{equation}
It can be verified that for any arbitrary value of $\lambda$, Eqs. \ref{eq4par} and  \ref{eq5per} are asymptotically, i.e., $h$ approaching infinity, correct   \cite{benavides2016brownian}; For  $\lambda\rightarrow\infty$  it replicates the behavior near a hard wall, and for  $\lambda\rightarrow 0$  the model correctly recovers the behavior near a free interface. This approach offers a unified theoretical framework encompassing both fluid-fluid and solid-fluid interfaces \cite{benavides2016brownian}.

\section{Experimental data analysis procedure}
\label{analysis}
By combining the digital image processing approaches with video microscopy, detailed and time-specific tracking of individual micro-particles within the suspension will be achieved. 
Since Perrin's groundbreaking image-based exploration of diffusion and Brownian motion nearly a century ago  \cite{perrin1920}, the quantitative evaluation of colloids has been employed for several phenomena, for example for investigating phase transitions in  2D and 3D systems, examining the impact of external fields on the dynamics of colloids, and directly quantifying the interaction between distinct pairs of colloidal microspheres \cite{crocker1994microscopic}. 
The application of digital video analysis on the captured microscopic images allows for the determination of the paths followed by individual micro-particles. This method also facilitates the study of how the micro-particle distribution changes over time.

A direct method to measure particle diffusion involves tracking the individual movement of colloids in sequences of images at set time intervals and visualizing the Brownian motion of the particles. 
The center of the particles are determined and the distances they travel over each time interval are measured. The mean square displacement (MSD) is then plotted against time to directly derive the diffusion coefficient.
Several studies have employed this principle. For instance, Crocker and Grier used digital video microscopy in conjunction with optical tweezers to investigate the hydrodynamic corrections of Brownian motion for two spheres in close proximity, and even the direct interactions for charged colloids \cite{crocker1994microscopic}. This approach achieved an experimental diffusion measurement accuracy of $\pm$1\%. It is noteworthy that the Brownian motion of particles allows for the estimation of Boltzmann's constant and Avogadro's number \cite{nakroshis2003measuring,salmon2002brownian}.
Park et al. applied optical serial-sectioning microscopy to extend the measurement of the displacements of Brownian particles in 3D \cite{park2005temperature}.

In the context of our study, a particle refers to a micro-particle that we trap at the oil-water or air-water interfaces using ST.  
Micro-particles can be made from various materials, including polymers, proteins, lipids, and metals, and can range in size from a few nanometers to a few micrometers. It is obvious that the specific properties of the micro-particles, such as their size, shape, and material, can significantly influence their behavior when trapped at fluid interfaces.
The coordinates of a particle refer to its position in space at a given time. In a 2D system, a particle's coordinates are usually given as $(x,y)$, where $x$ is the horizontal position and $y$ is the vertical position.   We record numerous videos of these micro-particles in motion, extract their coordinates at different times across multiple frames, and record their positions in each frame.
The resulting set of coordinates forms a trajectory that describes the particle's motion over time. By analyzing the trajectories, we can learn about the forces acting on the particles and their response to these forces.
For example, if a particle moves in a straight line, its coordinates will change linearly with time. If a particle is trapped in a harmonic potential well, its coordinates will oscillate around the minimum of the potential. By studying these patterns, we can gain insights into the nature of the forces acting on the particles and the properties of the fluid interfaces.

The MSD is a statistical measure used in the fields of physics, chemistry, and biology to quantify the spatial extent of micro-particle motion as a function of time. It is particularly useful in the study of random processes, such as Brownian motion or the diffusion of particles in a fluid.
The MSD of a micro-particle is calculated as the average of the squared differences in positions that a particle experiences over time. Mathematically, it is represented as:
\begin{equation}
{\rm{MSD}}(\Delta t) = \langle |r(t+\Delta t) - r(t)|^2\rangle,
\label{eq4}
\end{equation}
where $r(t)$ is the position of the micro-particle at time $t$, $\Delta t$ is the time lag, and  $\langle ...\rangle$ denotes a thermodynamic average over all time origins.
The first step to extract the MSD is calculating the square of displacements of the tracked particles  ($|r(t + \Delta t)$ - $r(t)|^2$). This is performed for all time points ($t$) and all-time lags ($\Delta t$). Then, their average is taken by summing up all the squared displacements and dividing by the number of time origins. 

We derive the mean MSD over an ensemble of particles. This gives an MSD value that represents the average behavior of all the particles in the system in a specific condition.  
The MSD provides insights into the nature of the micro-particle's motion. 
For example, in a log-log plot if the MSD is linearly proportional to the time lag $\Delta t$, i.e., MSD($\Delta t) \propto \Delta t$, it will indicate a diffusive motion, which is the characteristic of simple Brownian motion. 
On the other hand, if the MSD is proportional to $\Delta t^{2}$, i.e., MSD ($\Delta t) \propto \Delta t^{2}$, it will indicate a ballistic or directed motion, which is the characteristic of a micro-particle moving at a constant velocity that can be caused by flowing fluid. According to the above points, careful experimental design and data analysis are crucial for obtaining reliable MSD estimates.
We denote the trajectory of a Brownian particle (when the laser is switched off) by $r(t)$, which is characterized by its self-diffusion coefficient  $D$, as expressed by the Einstein-Smoluchowsky equation is $ \langle|{\bf r}(t+\Delta t)-{\bf r}(t)|^{2}\rangle=2dD\Delta t$, where $d$ represents the dimension of the trajectory data.  
Therefore, time-resolved particle trajectories obtained from digital video microscopy observations offer a straightforward method for determining the self-diffusion coefficients of colloidal micro-particles. As a standard application of the aforementioned imaging techniques, measurements of the diffusion coefficient serve as a quantitative verification of the accuracy of our particle-location and track-reconstruction methods. As an analytical instrument, such measurements supplement traditional light scattering techniques by allowing for direct measurements on polydisperse, inhomogeneous, strongly interacting, or highly dilute suspensions.

In our study, we first quantify the random motion of micro-particles at the interfaces of water, oil, and air using the  MSD. Building on this foundation, we then employ the Force Reconstruction via Maximum-likelihood-estimator Analysis (FORMA) algorithm \cite{perez2018high}. FORMA, by analyzing the displacements of these Brownian particles, allows us to infer the most likely forces that are resulted in the observed displacements. The algorithm utilizes a linear maximum-likelihood-estimator toward leveraging the statistical properties of Brownian motion and unraveling the underlying force landscape.  This powerful computational tool enables us to calculate both the conservative and non-conservative components of the estimate force field acting on microscopic systems, such as optical tweezers, thereby,  providing a comprehensive understanding of the interactions between the micro-particles and the surrounding medium. The FORMA algorithm has several advantages over other predated developed techniques; it is parameter-free, requires ten-fold less data, and executes orders-of-magnitude faster. By implementing the FORMA algorithm, we can enhance our understanding of the forces and make more precise predictions. This method is particularly beneficial in the realm of biophysics, where comprehending the forces acting on biological particles can yield insights into microscopic biological processes.

Unlike conventional force fields with known equilibrium positions, speckle patterns present a unique challenge due to their random nature and virtually infinite configuration space. Furthermore, speckle fields can exhibit non-conservative components, further complicating their analysis.  Nevertheless, FORMA has the possibility to be extended for effective analysis of complicated force fields, including the intricate random optical forces generated by speckle patterns.  
This is because that it estimates the force field directly from the observed trajectories of particles, bypassing the need to calculate the potential energy landscape.   

The acquisition of particle trajectories is obtained by monitoring and recording the motion of the particles across a sequence of images or videos. Upon obtaining the trajectories, the FORMA algorithm proceeds to estimate the force field  by solving the equation:
\begin{equation}
	F_{ext} = -\nabla U,
\end{equation}
In this equation, $	F_{ext}$ represents the force, $\nabla$ denotes the gradient operator, and $U$ signifies the potential energy. In systems characterized by non-conservative forces, where $U$ is unknown, FORMA directly estimates $	F_{ext}$.
The force field that maximizes the likelihood of the observed trajectories, given the force field and the noise level, is then selected as the estimated force field using the maximum-likelihood estimator. The final step involves the reconstruction of the estimated force field from the estimated forces. This reconstructed force field offers valuable insights into the forces acting on the particles and can be utilized to comprehend and predict their behavior.

\begin{figure*}[t!]
	\begin{center}
		\includegraphics[width=0.99\linewidth]{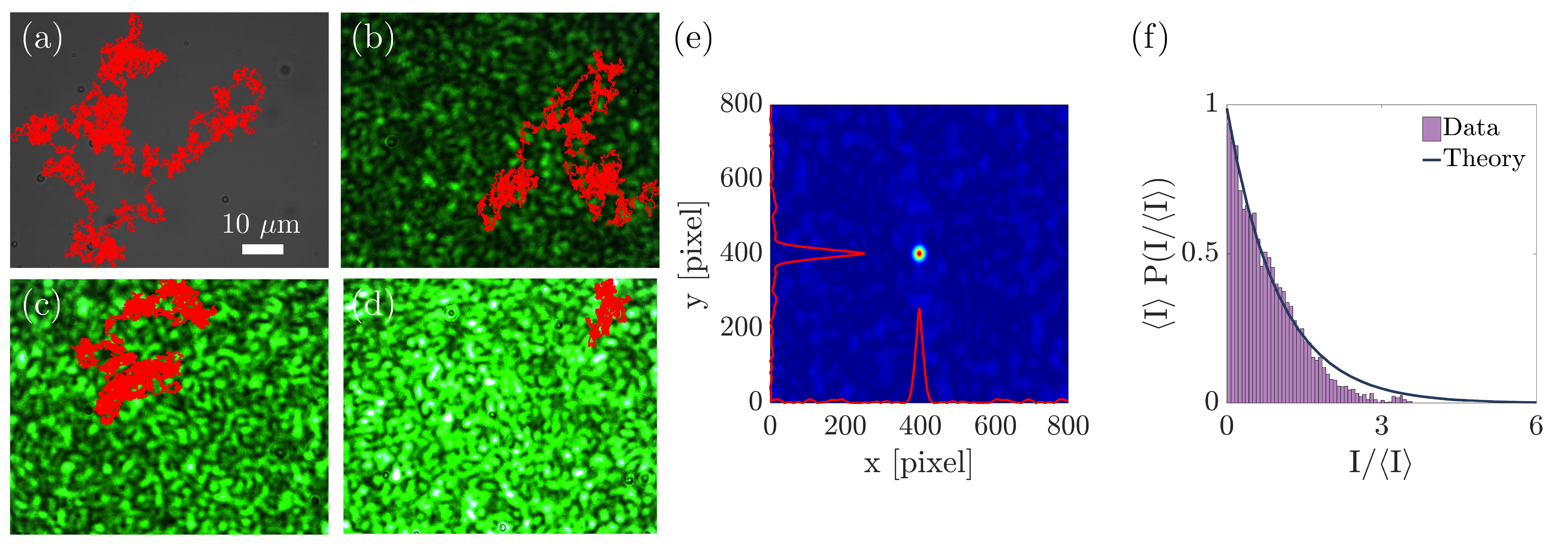}
		\caption{The experimental trajectories (red solid lines) of particles at water-oil interface in progressive confinement through increasing the average speckle intensity for the duration of 1000 s.  In the experiment, polar and non-polar Newtonian fluids are used to create a common interface between two fluids. The refractive index and the viscosity of distilled water are $n_m = 1.33$ and $\eta_w$ = 0.001 $\rm{Pa.s}$, respectively. The refractive index and the viscosity of the oil are $n_o = 1.37$ and $\eta_o$ = 0.122 $\rm{Pa.s}$, respectively. 
		The background represents the speckle field generated by the propagating of a solid-state laser light through a multimode fiber.  Trajectories of a PS micro-particle of the refractive index of $n_p = 1.59$ and diameter of $D_p = 3$ $ \mu$m  near the water-oil interface in the cases of (a) no speckle filed (4D$_0$$\Delta t$), (b) $\langle I\rangle = 5.46 $ $\mu$W/$\mu \rm{m}^2$, (c) $\langle I\rangle = 3.24 $ $\mu$W/$\mu \rm{m}^2$, and (d) $\langle I\rangle = 1.08 $ $\mu$W/$\mu \rm{m}^2$. (e) Normalized spatial autocorrelation function, which characterizes the average speckle grain size as the full width at half maximum (FHWM) of the autocorrelation along the $x$ and $y$ axes (solid red lines).
		The average speckle grain size for our experiments is $4.46 \pm 0.20$ $\mu$m, and a pixel demonstrates $0.109 \mu$m at the sample plane. (f) The intensity distribution of the speckle pattern at the sample plane and negative exponential function fitting. The violaceous bars indicate the experimental data and the navy blue solid line demonstrates the fitted theoretical probability density function, which follows the negative exponential distribution.}
		\label{Figure03}
	\end{center}
\end{figure*}

\section{Results and discussion}
\label{results}
In our exploration of micro-particle manipulation using a speckle field at a water-oil interface, we begin by examining the movement patterns of a sample composed of PS micro-particles with a diameter of 3 $\mu$m and a refractive index of 1.59 when exposed to a laser field with random distribution. The output end of a multimode fiber is introduced into the specimen chamber to generate and apply ST to the dispersed particles. The average intensity of the speckle grains is regulated by adjusting the laser light intensity, and the grain size can be altered by meticulously adjusting the distance between the fiber end and the sample.
Figure \ref{Figure03} illustrates the trajectories of a typical micro-particle in a water-oil interface under varying speckle fields with low to high average speckle intensity (panels (a) to (d) in \ref{Figure03}) for a duration of 1000 s. 
When the average speckle intensity is relatively zero, i.e., the ST is switched off, the particle has freely diffusing behavior (Fig. \ref{Figure03}(a)).
When the ST is on, the trajectory of the PS micro-particle reveals the relocation of it toward the bright areas of the speckle pattern. Under weaker optical forces, for example in  Figs. \ref{Figure03}(b) with $\langle I\rangle = 1.08 $ $\mu$W/$\mu \rm{m}^2$, the particle stays very briefly-kept in one of the bright speckle grains. The particles exhibit nearly free diffusion but still the interaction with the speckle field confines them to some small extent.
In line with their Brownian motion, the particles also have the ability to move within the speckle grain.
 By increasing the average laser intensity to $\langle I\rangle = 3.24 $ $\mu$W/$\mu \rm{m}^2$, however, the micro-particles remain metastably trapped. Yet, they can make transitions between adjacent grains (Fig. \ref{Figure03}(c)), as their trajectories show. 
Further increasing the average intensity to $\langle I\rangle = 5.46 $ $\mu$W/$\mu \rm{m}^2$ (Fig. \ref{Figure03}(d)), which causes to stronger optical forces, and in turn forces the micro-particle remaining trapped in one or a very limited number of the bright speckle grains for several minutes.  The optical force that acts on a moving particle changes with a characteristic time scale within the speckle field, which is inversely related to the average speckle intensity in the first-order approximation \cite{volpe2014brownian}. 
The net motion of a Brownian particle in a static speckle field is the outcome of random thermal forces and deterministic optical forces. In these experiments, the optical forces compete with the Brownian motions. 
Figure \ref{Figure03}(e) depicts the normalized spatial autocorrelation function of the speckle pattern used in the experiment.
 Determination of the speckle grain size is crucial for analyzing the interaction between the light field and the micro-particles. The average speckle grain size is primarily determined by the diffraction of light and is calculated by measuring the FWHM of the autocorrelation function \cite{goodman1976some,goodman2007speckle}. 
It is done by fitting  Gaussian functions along   $x$ and $y$ axes, which are shown as overlaid plots in Fig. \ref{Figure03}(e).
 The autocorrelation function of Fig. \ref{Figure03}(e) estimates the grain size for our experiments as 4.46 $\pm$ 0.20 $\mu$m.  
Figure \ref{Figure03}(f) illustrates both the theoretical  (solid navy-blue line) and experimental (purple bars) probability density function of a standard speckle pattern at the sample plane. The probability density function of the speckle pattern's intensity distribution adheres to the negative exponential distribution, denoted as $\frac{1}{\langle I\rangle} \exp{(-\frac{I}{\langle I\rangle})}$ \cite{goodman2007speckle,bender2018customizing}. The agreement between the theoretical predictions and the experimental results of the distributions of the speckle pattern intensities is evident in Figure \ref{Figure03}(f). 

\begin{figure*}[t!]
	\begin{center}
		\includegraphics[width=0.99\linewidth]{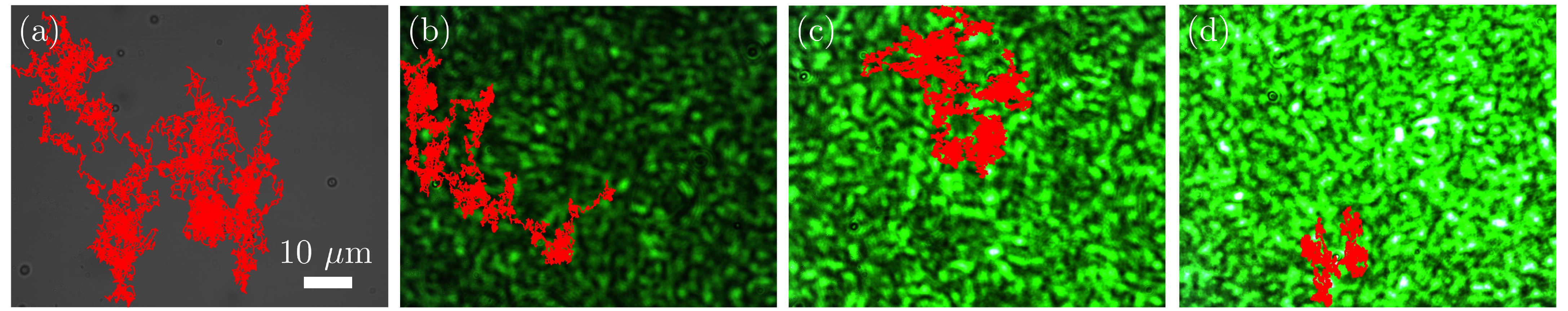}
		\caption{The experimental trajectories (red solid lines) of PS micro-particles in progressive confinement through increasing the average speckle intensity for the duration of 1000 s.  In this experiment, water and air are used to create a common interface and it is performed through a hanging droplet mechanism. The refractive index and the viscosity of distilled water are $n_m = 1.33$ and  $\eta_w$ = 0.001 $\rm{Pa.s}$, respectively, and the refractive index and the viscosity of air are $n_a = 1.00$ and $\eta_a$ = $1.81 \times 10^{-5}$ $\rm{Pa.s}$, respectively.  The background represents the generated speckle field.  Trajectories of a PS micro-particle of the refractive index of $n_p = 1.59$ and diameter of $D_p = 3$ $ \mu$m  near the water-air interface are overlaid. (a) No speckle filed (4D$_0$$\Delta t$), (b) $\langle I\rangle = 5.46 $ $\mu$W/$\mu \rm{m}^2$, (c) $\langle I\rangle = 3.24 $ $\mu$W/$\mu \rm{m}^2$, and (d) $\langle I\rangle = 1.08 $ $\mu$W/$\mu \rm{m}^2$.}
		\label{Figure04}
	\end{center}
\end{figure*}

Figures \ref{Figure04}(a-d) depict the path trajectories of representative micro-particles at water-air interface, subjected to speckle fields of progressing intensities: $\langle I\rangle = 0$ (Fig. \ref{Figure04}(a)),  $\langle I\rangle = 1.08 $ $\mu$W/$\mu \rm{m}^2$ (Fig. \ref{Figure04}(b)), $\langle I\rangle = 3.24 $ $\mu$W/$\mu \rm{m}^2$ (Fig. \ref{Figure04}(c)),  and  $\langle I\rangle = 5.46 $ $\mu$W/$\mu \rm{m}^2$ (Fig. \ref{Figure04}(d)). 
Similar to the case of the water-oil interface, the micro-particles under exposure to ST at the interface of water to air follow alike behavior; in the no-ST case the particle essentially diffuses freely, and as the average intensity increases the particle is further confined in a further limited number of speckle grains. In Fig. \ref{Figure04}(d) the micro-particle remains confined in one of the bright speckle grains for an extended duration of several minutes.

\begin{figure*}[t!]
	\begin{center}
		\includegraphics[width=0.9\linewidth]{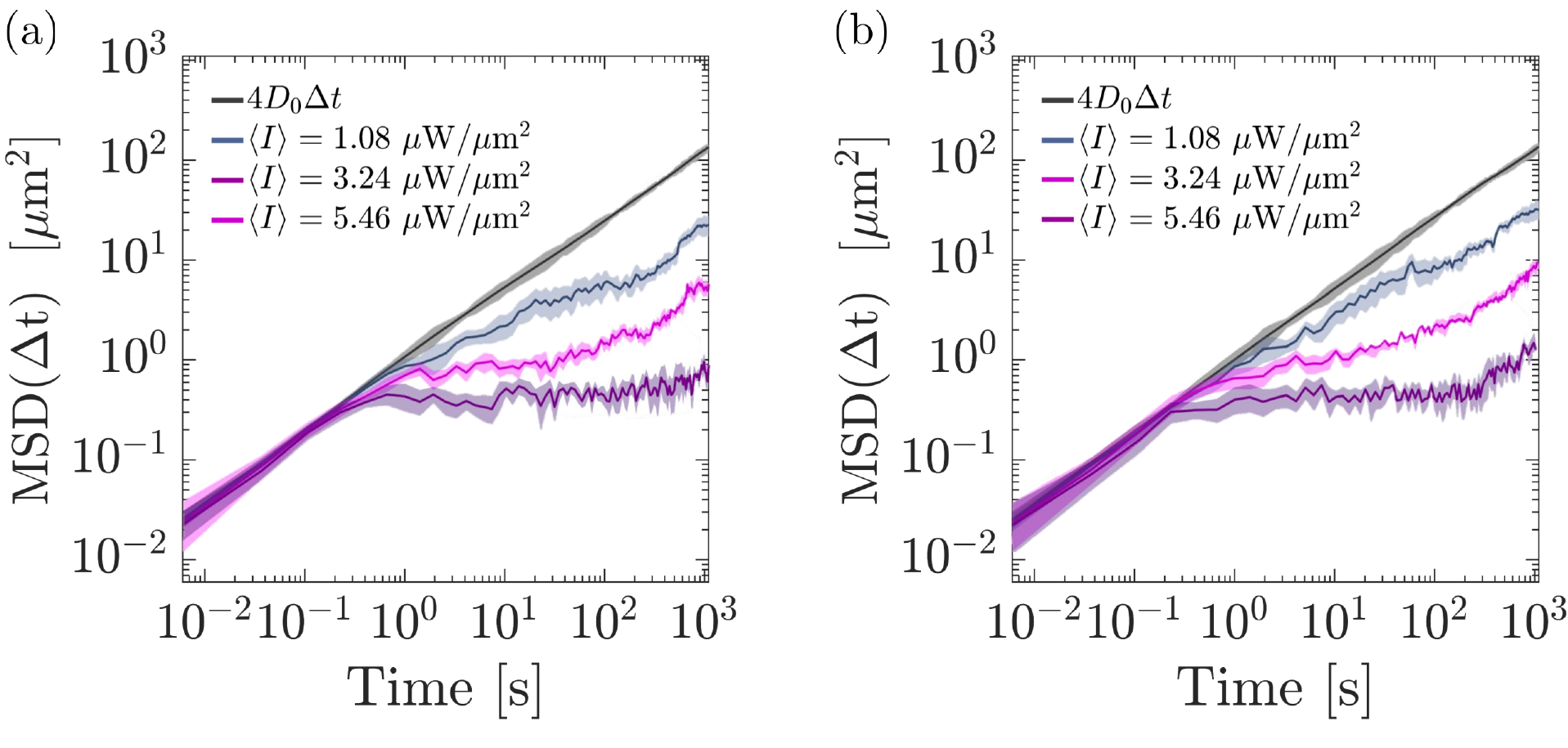}
		\caption{Mean square displacement (MSD) of Brownian particles on a logarithmic scale as a function of time in increasing speckle field intensity. (a) Water-oil interface; (b) Water-air interface. The particle initial position is randomly chosen. The data presented in each picture is the average of more than 1000 distinct trajectories during a period of 1000 s.  The average speckle intensity is increased from 1.08 to 5.46  $\mu$W/$\mu \rm{m}^2$. The black line represents the  Einstein's free diffusion law. }
		\label{Figure05}
	\end{center}
\end{figure*}

In order to illustrate the ST resulted confinement of the micro-particles in a more quantitative fashion we calculate their MSDs. The results are shown in Fig. \ref{Figure05}. The MSD characterizes the average spatial movement of a micro-particle within a static speckle field. It is calculated by averaging the squared displacements of the micro-particle's position over time and across the associated ensemble of micro-particles. Equation~\ref{eq4} formally defines the MSD.  
The MSDs of a PS micro-particle in the water-oil and water-air interface under static speckle field illumination are shown in Figs. \ref{Figure05}(a)  and \ref{Figure05}(b), respectively. The results are presented for three different average speckle intensities of $\langle I\rangle$=1.08, 3.24, and 5.46  $\mu$W/$\mu \rm{m}^2$.
The MSD of control experiment, when the speckle field is switched off, is also shown. As expected, in both interfaces it follows a linear behavior with the slope of $4 D_0$, $D_0$ being the diffusion coefficient. 
Each MSD curve is derived by averaging over 1000 different trajectories for a time duration of 1000 s and with the acquisition frame rate of 15 fps.   
Brownian motion, the random and ceaseless dance of microscopic particles suspended in a fluid, exhibits dependence on the surrounding environment. Therefore, the Brownian motion at the water-air and water-oil interfaces demonstrated slightly different behavior.
 The key to understand this disparity lies in the interplay between viscosity and interfacial tension. Water boasts a significantly lower viscosity compared to most oils, and, therefore, in the lower-viscosity water environment, particularly at the water-air interface, the micro-particles encounter less friction, allowing for more pronounced Brownian motion. Furthermore, in the water-air interface, the interfacial tension is higher compared to the water-oil interface, and creates a more stable environment for the micro-particles. This stability allows them to undergo more significant Brownian motion without being hindered by interfacial fluctuations \cite{wang2020brownian}. 
In our experiment, the confluence of these two factors, i.e.,  the lower viscosity of water and the higher interfacial tension at the water-air interface, leads to a more prominent display of Brownian motion for the micro-particles at the water-air interface compared to the water-oil interface.

\begin{figure*}[t!]
	\begin{center}
		\includegraphics[width=0.8\linewidth]{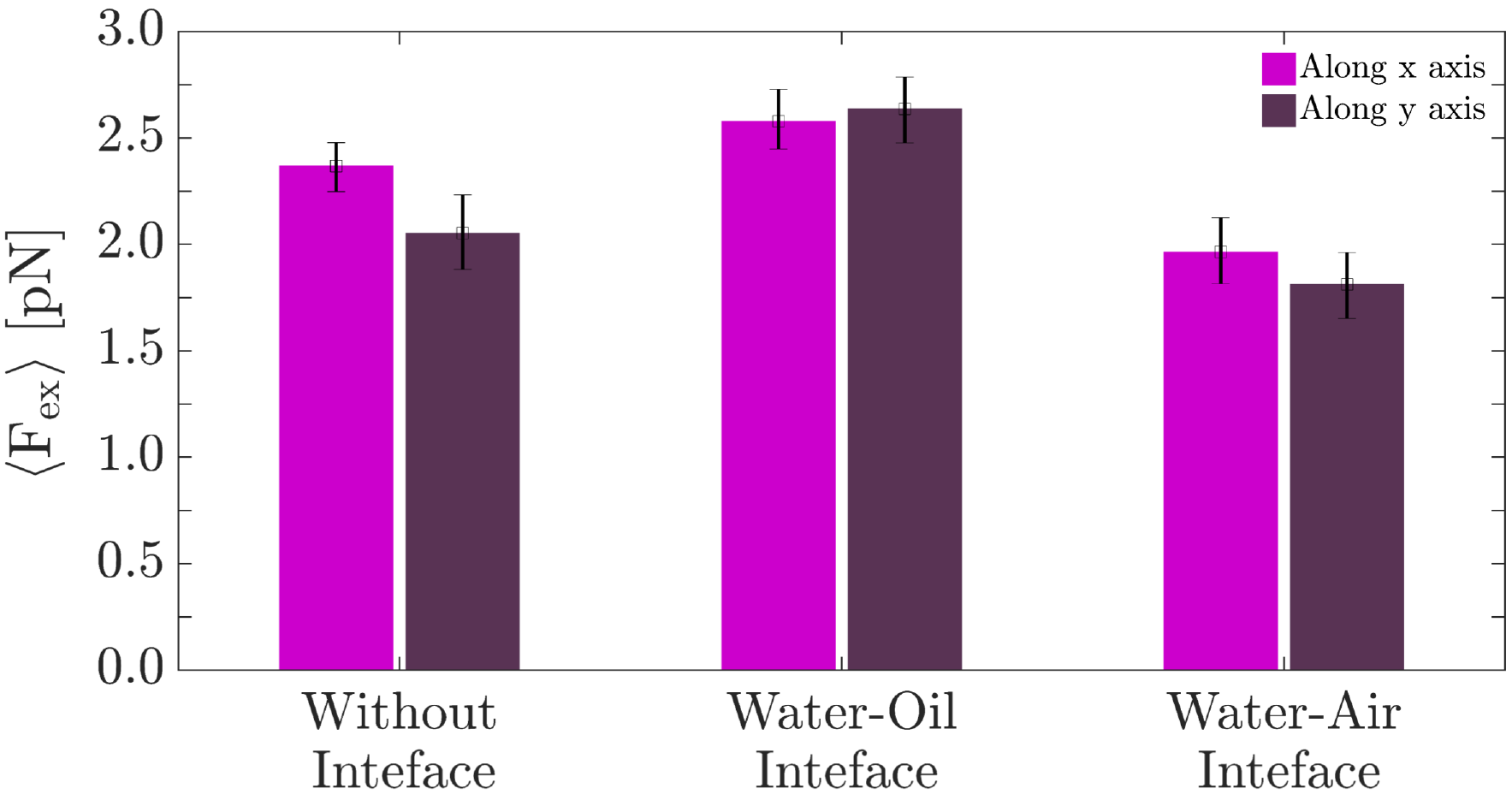}
		\caption{The average forces ($\rm{\langle F_{ex} \rangle}$) exerted on micro-particles at the water-oil and water-air interface and in an interfaceless state. The graph provides a FORMA algorithm detailed view of the dynamic interactions and transitions experienced by the micro-particles under speckle field conditions.}
		\label{Figure06}
	\end{center}
\end{figure*}

When the SP is switched on, the relationship between MSD and time ($\Delta t$) reveals the interplay between optical and thermal forces. 
At low forces and short timescales, the MSD in both water-oil and water-air interfaces exhibits a almost linear dependence on $\Delta t$ (MSD $ \sim4D_{\rm{SE}}\Delta t$), signifying free Brownian motion with a diffusion coefficient of D$_{SE}$ (Stokes-Einstein diffusion coefficient). 
As forces increase or timescales decrease, the system transits to  subdiffusive regime, characterized by MSD $\propto\Delta t^{\beta} $ with $\beta < 1$. 
This indicates hindered micro-particle movement with a reduced effective diffusion coefficient (D$_{eff}$). At very long timescales, even under strong forces, the MSD may eventually recover linear behavior ($\beta \approx$ 1) with a lower D$_{eff}$, reflecting a potentially caged diffusion within a limited region.
For higher speckle intensities, i.e., with stronger forces, the micro-particles after an initial transient diffusion get trapped within speckle grains. At even higher forces, the micro-particles become completely confined within a single speckle grain, avoiding even inter-grain movements.
This observed transition from free diffusion to subdiffusion and potential trapping behavior happens for all the examined micro-particles under ST. 

Furthermore, in order to measure the optical forces acting on the particles at the interfaces and compare them with the case of no-interface case,  we obtain the microscopic quantitative force fields through force reconstruction via the maximum-likelihood-estimator analysis (FORMA) algorithm \cite{perez2018high}.
 In FORMA, the forces are retrieved by the analysis of the particles trajectories. The optical force fields generated by ST are larger extended force fields and the equilibrium positions are not known \emph{a priori} due to their random appearance and include non-conservative components. FORMA estimates accurately the conservative and non-conservative components of the force field simultaneously with important advantages over the common methods that obtain the forces by analyzing their influence on the particles Brownian motion. This method is more accurate, does not need calibration fitting parameters, executes faster, and requires fewer data \cite{perez2018high}. 
The estimated exerted force along the $x$ and $y$ axes for no-interface, water-oil interface, and water-air interface are depicted in Fig. \ref{Figure06}. For these plots we employ long trajectories for the force measurements in FORMA, therefore, it ascertains reaching equilibrium and thoroughly exploring of the particles in the field of view. 
Water molecules are attracted to each other due to polar interactions. This attraction creates a tension at the surface of the water, like a thin elastic sheet.  On the other hand, oil does not have the same strong attraction between its molecules, so the surface tension of oil is much weaker than that of water, as we explained earlier.
Air, being a gas, has very weak intermolecular forces, so the surface tension of air is negligible.
The micro-particle at the water-oil interface would experience a force due to the water surface tension pulling it down (into the water) and a force due to the oil surface tension pulling it up (into the oil). Since water has a higher surface tension than oil, the net force on the particle would be to pull it into water. This would be a stronger force than the force experienced at the water-air interface, where the air surface tension is negligible.
Figure \ref{Figure06} also includes a section depicting a micro-particle suspended in a volume of pure distilled water (without interface).  This serves to illustrate the concept of surface tension even in the absence of an external boundary, like air or oil. Water molecules at the surface experience a net inward pull from surrounding water molecules due to hydrogen bonding \cite{silverstein1998real,binks2017colloidal}. This inward force is the essence of surface tension.
Consistently,  Fig. \ref{Figure06} likely shows that a micro-particle experiences a stronger net force when it is at the water-oil interface compared to the water-air interface.

\section{Conclusion}
\label{conc}
In conclusion, we have empirically validated an innovative method for optically manipulating micro-particles at fluid-fluid interfaces utilizing static speckle fields. While a meticulously pre-designed periodic potential or holographic optical tweezers tailored for a specific application in bulks, the present method, due to its quasi-2D nature, broadens the toolkit available to scientists and engineers for executing complex micro-manipulation tasks at interfaces and surfaces. Our strategy can be also re-scaled to meet the high throughput or sensitivity demands of microfluidics by enhancing interface manipulation with a low laser power.  Speckle tweezers possess further benefits over holographic optical tweezers, such as inherent consistency to environmental noise and optical aberrations, and easy implementation. The method has been applied for the collective trapping of polystyrene micro-particles at water-oil and water-air interfaces. 
The movement of the micro-particles was analyzed using their trajectory and mean square displacement as a measure. The MSD results showed that particles underwent subdiffusive motion, i.e.,  briefly trapped or experienced motion hindering.   This hindered motion was more pronounced at the water-oil interface compared to the water-air interface.
Force reconstruction via maximum-likelihood-estimator analysis algorithm was also used for analyzing complex force fields and to estimate the forces acting on the micro-particles. The results suggest that the water-oil interface exerts a stronger net force on the micro-particle compared to the water-air interface, which is in consistence with the trajectory tracking. This difference is attributed to the interplay between surface tension forces at the interfaces. 
Speckle tweezers has the potential for collective motion control of mixtures of micro-particles of various shape, size, and refractive index, which is a required task in many significant researches and applications.

\section*{Competing interests}
The authors declare no competing interests.
\section*{Data availability}
The datasets used or analyzed during the current study available from the corresponding author on reasonable request.
\section*{Author contributions}
A.M. conceived and supervised the project. R.J. carried out the experiments and analyzed the data. All authors interpreted and discussed the results and contributed to the writing and reviewing the manuscript. 
\section*{Acknowledgment}
S.K.P.V. acknowledges the Department of Biotechnology (DBT), Ministry of Science and Technology, Govt. of India, New Delhi, and Rathinam College of Arts and Science, Department of Physics, under the DBT Star College Scheme for the collaborative research support.

\bibliography{sample.bib}

\end{document}